# Topology analysis and *URANS*-predicted flow field features of multiple strongly impinging jets-in-crossflow in a cylindrical duct

James D. Holdeman[1], Evgeniy V. Kartaev[2,*], John F. Foss[3], Marat G. Ktalkherman[2,**]

[1]retired. National Aeronautics and Space Administration, John H. Glenn Research Center, Cleveland, OH 44135 USA.
E-mail: jjdholdeman@aol.com

[2]SB RAS, Khristianovich Institute of Theoretical and Applied Mechanics, Novosibirsk 630090 Russia

[*]E-mail: kartayev@mail.ru

[3]Professor Emeritus, Michigan State University, East Lansing, MI 48824 USA
E-mail: foss@msu.edu

[**]E-mail: ktalkhermanm@mail.ru

## Abstract

The flow field studied was eight strongly impinging, radially injected jets, into a non-swirling mainstream flow in a cylindrical duct. Our previous paper (*Heat Mass Transf. (2020) 56:2285–2302*), showed that asymmetry in the solution is very likely to be an indication of the flow unsteadiness. Also, if the flow is asymmetric, even a time dependent flow solver would not represent the flow correctly if the computational domain is less than the complete cylindrical duct. Thus, the current numerical and topological study was conducted in a 360° (complete cylindrical duct) computational domain using a time dependent Unsteady Reynolds-Averaged Navier-Stokes *URANS* code.

Results of this study confirm that the flow of strongly impinging jets is unsteady and asymmetric after the establishment of a regular flow pattern. The asymmetry appears to be cyclic at a frequency of < 2 Hz. Also, the computational fluid dynamics (*CFD*) results agree with the rules of topological analysis from applied mathematics.

## Nomenclature

| | | |
|---|---|---|
| $C$ | $=(S/H)(J^{1/2})$ *(rectangular duct)* $=\pi(2^{1/2}/n)(J^{1/2})$ *(cylindrical duct)* | parameter of indication of the jet penetration |
| $D$ | | diameter of cylindrical duct, [mm] |
| $d$ | | orifice diameter, [mm] |
| $DR$ | $=\rho_j/\rho_m$ | jet-to-mainstream density ratio |
| $H$ | | rectangular duct height, [mm] <br> for cylindrical duct, $H=D/2$ |
| $h$ | | radial jet penetration depth, [mm] |
| $J$ | $=(\rho_j V_j^2)/(\rho_m U_m^2)$ <br> $=(DR)R^2$ | jet-to-mainstream momentum-flux ratio |
| $MR$ | $=(DR)^{1/2} J^{1/2}(n)(d/D)^2$ | jet-to-mainstream mass-flow ratio |
| $n$ | | number of orifices |
| $R$ | $=V_j/U_m=J^{1/2}/DR^{1/2}$ | jet-to-mainstream velocity ratio |
| $S$ | | spacing between compatible locations of adjacent orifices e.g. midplane-to-midplane, [mm] |
| $T$ | | gas temperature, [K] |
| $U$ | | axial velocity of gas flow, [m/s] |
| $V$ | | radial velocity of gas flow, [m/s] |
| $x$ | | axial coordinate <br> ($x>0$ – downstream of *JIP*) <br> ($x<0$ – upstream of *JIP*), [mm] |
| $r$ | | radial coordinate, [mm] |

## Greek symbols

| | |
|---|---|
| $\kappa$ | turbulence kinetic energy, [J/kg] |
| $\varepsilon$ | turbulence dissipation rate, [m$^2$/s$^3$] |
| $\theta$ | azimuthal coordinate |
| $\rho$ | gas density, [kg/m$^3$] |
| $\chi$ | topology parameter, Chi |

## Subscripts

| | |
|---|---|
| *j* | Jet |
| *m (c)* | mainstream (crossflow) |

## Acronyms

| | |
|---|---|
| *CFD* | *Computational Fluid Dynamics* |
| *FOU* | *First-order Upwind discretization method* |
| GTE | *Gas Turbine Engines* |
| JIC | *Jet(s)-in-Crossflow* |
| JICF | *=JIC* |
| JIP | *Jet Injection Plane* |
| LDA | *Laser Doppler Anemometry* |
| LES | *Large Eddy Simulation* |
| PIV | *Particle Image Velocimetry* |
| RANS | *Steady-state Reynolds-Averaged Navier-Stokes* |
| RFZ | *Recirculating Flow Zone* |
| RKE | *Realizable $\kappa$-$\varepsilon$ model* |
| RQL | *Rich-burn/Quick-mix/Lean-burn* |
| SIMPLE | *Semi-Implicit Method for Pressure Linked Equations* |
| SOU | *Second-Order Upwind discretization method* |
| TA | *Topology analysis* |
| TDR | *Turbulence Dissipation Rate* |
| TKE | *Turbulence Kinetic Energy* |
| URANS | *Unsteady RANS* |
| V/STOL | *Vertical/Short Take Off and Landing* |

# 1 Introduction

## 1.1. Jets-in-crossflow (*JIC*) in general

Mixing of jets-in-crossflow (*JIC* (aka *JICF*)) is of importance in many technological applications, including chimney exhaust plumes, flows associated with vertical/short takeoff and landing (*V/STOL*) aircraft (e.g. the Harrier and Yak-38), mixing in combustion chambers for gas turbine engines (*GTE*'s), and mixing in furnaces and in chemical reactors. Comprehensive reviews of subsonic, single and multiple, confined and unconfined, jets are presented in [1-8]. There are many papers on *JIC*'s that post-date the summaries in [1-5], including experimental reacting and non-reacting flow studies in a cylindrical duct [9-17] and *JIC* studies using the spreadsheet version of the NASA empirical model for *JIC*'s in a rectangular duct [18-25].

*JIC*'s have been studied in cylindrical, rectangular, and annular geometries motivated by mixing in conventional, reverse-flow, and Rich-burn/Quick-mix/Lean-burn (*RQL*) combustors for *GTE*'s (e.g.[2-4&9-27]). The design and optimisation of rapid-mix reagent units or quench zones in chemical reactors [e.g. 28&29] requires understanding of the *JIC* formed by multiple jets injected into a crossflow in a cylindrical duct, creating an upstream recirculation flow zone. Ktalkherman et al.[30] proposed the *h/D* parameter for generalization of the acquired experimental data in order to evaluate mixing effectiveness of multiple jets that interact strongly with crossflow. This approach was then applied for analyzing the parameters of *JIC* pyrolysis reactor [31] and those of quenching process in the plasmachemical reactor. Recent papers by Kartaev et al.[32] and Kartaev&Holdeman [33] were devoted to experiments and analysis of interacting impinging jets and crossflow and their mixing. Kartaev&Holdeman [33] showed that the results of the study for a counter flow jet-in-crossflow (*JIC)* mixer are consistent with, and extend, the experimental and numerical results published earlier.

The experimental *JIC* results reported recently by Thong et al.[34] are of special interest as they provide experimental evidence that the impinging *JIC* flow in a cylindrical *JIC* mixer is unsteady at high momentum-flux ratios. Although Kartaev et al.[32] didn't suspect unsteadiness from their experimental results, they looked for it based on the results reported in [34], and an appearance of asymmetry of developed *JIC* flow and a further establishment of flow unsteadiness were evident in Unsteady Reynolds-Averaged Navier-Stokes (*URANS*) simulation results shown in [32] and [33].

The reports and papers by Foss [35&36] and Foss et al.[37] are also of special interest as they focus on the application of topological analysis from applied mathematics to *JIC*'s.

Although jet-to-mainstream momentum-flux ratio *J* was shown in e.g.[2-4] and elsewhere (including Sequences 2-4 in the slideshow (Appendix A) in [19]) to be the most important flow variable, the jet-to-mainstream mass-flow ratio *MR* is also important, as *MR/*(*MR*+1) is the dimensionless scalar indicating complete *JIC* mixing.

Note that flux variables, such as *J*, do not need knowledge of the area ratio, but flow variables, such as *MR*, do. From results obtained in a rectangular duct, Holdeman [2] proposed the parameter *C* as an indicator of the radial penetration of multiple jets where $C=(S/H)(J^{1/2})$ and *C* is directly proportional to the orifice spacing divided by rectangular duct height (*S/H*) and the square root of *J*. Note that neither the orifice diameter *d* nor the jet-to-mainstream density ratio *DR* appear explicitly in *C*. Also *S/H*>1 was never investigated in the experiment reported in [2] so *C* should not be greater than square root of *J*. For a cylindrical duct Holdeman [2] assumed that *S/H* should be applied at the radius that divides the cylindrical duct into equal-area annular and cylindrical parts. Thus, $(S/H)_{can}=\pi(2^{1/2})/n$ and $C=\pi(2J)^{1/2}/n$, wherein *C* is directly proportional to the square root of *J* and inversely proportional to the number of orifices *n*. The analogy between rectangular and cylindrical ducts becomes increasingly tenuous as *n* decreases. The relation $C=\pi(2J)^{1/2}/n$ has been used for *n*=6 and seems to apply for *n*=4, but should definitely not be used if *n*<4 as the flow in a cylindrical duct for *n*<4 would be expected to be a lot different than that in a rectangular duct at the same spacing and also $S/H_{can}$>>1 for *n*<4.

**1.2 Flowfield and objectives of this study**

The flowfield studied was eight strongly impinging, radially injected jets into a non-swirling mainstream flow in a cylindrical duct. The computational domain of this study is shown in Fig.1 in both a radial-axial centerplane and a radial-azimuthal plane.The eight jets issued to the cylindrical duct at the orifice center *x/D*=0 so distances upstream and downstream are negative and positive, respectively from the center of the jets (*JIP*). As can be seen in Fig.1b, the 8 jets are numbered clockwise from the 270° position (compass west). The centerplane for Jet1 is at the starting location and jets are centered at 45° intervals. In the figure also the radial-axial plane mid-spaced between adjacent jets 1&2 (and jets 5&6) is designated as midplane 1.5-5.5. In Fig.1a predicted temperature profiles in the horizontal centerplane (a radial-axial plane through the centers of opposed jets 1&5) are shown for the case of non-impinging *JIC* ($J^{1/2}$=4.9 (*C*=2.7)).

An important feature of radially injected *JIC*'s in a cylindrical duct is that the mixing region associated with each jet decreases in width as the duct centerline is approached from the perimeter. The effect of this is evident in Figs.10a&b in Vranos et al.[38].

This paper is a spinoff from our paper published previously in [33]. The cyclic analysis is not in [33]. In fact, cycling is time dependent, and thus cannot be estimated from results at a fixed time as reported in [33]. Results for the farthest upstream penetration of the counter flow jet were not included in [33]. Also, midplanes for $J^{1/2}$=39.6 were also not in [33] nor were the pressure data to confirm our cycling frequency estimates.

The (unconventional) Topology Analysis (*TA*) from applied mathematics was added to verify the integrity of our computational results because there are no experimental results in the open literature with which to compare our calculations for strongly impinging jets.

Specific objectives of the current study are:

1) Investigate results for strongly impinging jets from the jet injection plane (*JIP*) to the farthest upstream penetration of the counter-flow jet.
2) Investigate results for asymmetric impinging jets after the commencement of 'flapping' (the term from [34]).
3) Show that the asymmetry appears to be cyclic.
4) Verify the integrity of the *CFD* results by comparing them with the rules of topological analysis.

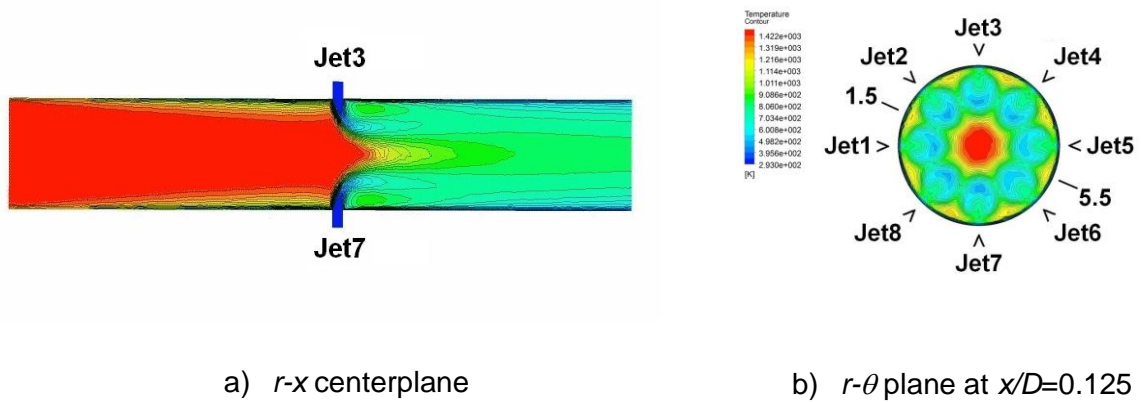

a)  *r-x* centerplane            b)  *r-θ* plane at *x/D*=0.125

**Fig. 1** *URANS* temperature plots at the side(a) and the end (b) views for 360° calculation domain (extracted from [33]) with jet #'s and locations shown; $J^{1/2}$=4.9, *n*=8 (*C*=2.7), *DR*=5 (*MR*=0.78), *t*=3.000 sec

## 2 Details of *JIC* flow conditions and numerical simulation
## 2.1 Numerical model

As the requirements of the *RANS* and *URANS* models for computational resources and calculation time are modest compared to *LES* and Direct Numerical Simulation, the current simulation was carried out with *RANS* and *URANS* codes.

The equation set was closed using a two equation $\kappa$-$\varepsilon$ model. This numerical approach is based on consideration of turbulence kinetic energy $\kappa$ and turbulence dissipation rate $\varepsilon$. Commercial *CFD* code Fluent 15.0 [39] was applied, adopting *realizable* $\kappa$-$\varepsilon$ model and Favre-averaging [40] and used in both *RANS*&*URANS* modes.

The model default parameters in [39] were specified. In particular, turbulence kinetic energy $\kappa$ (*TKE*) Prandtl number was equal to 1.0 and turbulence dissipation rate $\varepsilon$ (*TDR*) Prandtl number = 1.2.

It was assumed that the solution convergence was reached when the residuals were constant and reduced for all *RANS* equations by at least three orders of magnitude, and by 4-5 orders for the energy equation.

## 2.2 Discretization and gridding

For all governing equations (continuity, momentum, energy, *TKE*, and *TDR* equations) a first-order upwind discretization method (*FOU*) was used for the algebraic presentation of the convection terms whilst the diffusion terms are central-differenced and are always second-order accurate [39]. Diffusion terms and velocity derivatives in the governing set of equations evaluated using default Least-Square cell-based gradient method most suitable for computing on irregular (skewed) unstructured meshes.

The unstructured grid with total number of computational volumes of 395730 was chosen for numerical calculations in the cylindrical duct domain of interest. Grids were refined in the vicinity of the orifices. The minimum cell size near the orifices was about 9% of *d* (.84% of internal duct diameter *D*), with a cell growth rate of 1.10. The maximum size was 7.8% of *D*. The minimum orthogonal quality was about 0.243. The grid density of the computational domain was approximately the same as in Kartaev et al.[32].

It is known that all numerical methods introduce numerical (false) diffusion in modelling various kinds of fluids, McGuirk et al.[41] and Raithby [42]. The *FOU* scheme adopted is the least (first-order) accurate on the false diffusion, therefore numerical solutions obtained using *FOU* should be assessed from the point of impact of the false diffusion, in particular for the case of turbulent recirculating flows [41]. Although second-order upwind discretization scheme (*SOU*) is more accurate at specific flow conditions in comparison with the *FOU*, application of *SOU* and higher-order dicretization schemes could exhibit spurious flow fluctuations (wiggles) due to the introduction of a downstream/upstream cell in these schemes. These fluctuations might be confused with flow instabilities associated with an interaction of transversely injected jets and crossflow in the cylindrical duct, i.e. in the regions with the strong velocity gradients. It is of special importance in the *JIC* stagnation regions of unstable equilibrium nearby the point of jets' impingement. Thus, in convection dominated flows, applying the *FOU* constitutes heavy handed *JIC* flow stabilization at the cost of an increased numerical diffusion.

In the case of *FOU* the simple analysis is available to estimate the false diffusion in the terms of cell (grid) Peclet number $Pe_\Delta$. The analysis of cell Peclet number estimates in the computational sub-domains of interest for the working grid was conducted for the steady-state case of impinging *JIC* ($J^{1/2}$=9.9 (*C*=5.5, *MR*=1.56)). In Figs.11&12 of [33] there are shown the centerplane & radial-transverse plane distributions of temperature and streamlines for this case. The results of analysis of Peclet number profiles show that in convection dominated regions ($Pe_\Delta$>5) false diffusion is negligible in both axial and radial directions whereas within the recirculating flow zone (*RFZ*) the false diffusion error is evaluated to be lower than 4-5% [42] and doesn't substantially influence the *JIC* flowfield pattern.

In [42] it is stated that accuracy of upwind differencing (*FOU* herein) would deteriorate in the case of transient convection. It was found that large errors occur for all but very small frequencies. Our estimates based on [42] show that *FOU* scheme in the case of cyclic *JIC* motion of ~2 Hz results in less than 1% false diffusion error.

## 2.3 Flow and boundary conditions

Air was considered incompressible for both the mainstream flow and the injected jets. Standard wall functions were applied. Atmospheric pressure of the cylindrical duct is zero-referenced. Preliminary numerical tests in which the air specific heat, thermal conductivity and molecular viscosity were set as temperature-dependent variables demonstrated almost identical turbulent flow fields compared with those obtained when the thermal properties of the air were treated as constant. This is because at high Reynolds jet flows molecular transport processes are considered to be negligible compared with turbulent transport. To reduce computational cost, in further numerical calculations, thermal properties of the air were kept constant. The air temperature at the inlet boundary of the mainstream was $T_m$=1473K and the inlet temperature of the jets was $T_j$=293K for an approximate jet-to-mainstream density ratio of 5. In general, the velocity ratio $R=J^{1/2}/(DR)^{1/2}$ where $DR$ is the density ratio $=\rho_j/\rho_m$. Thus, the velocity ratio in [32&33] was less than half (45%) of $J^{1/2}$.

Steady-state boundary conditions were imposed for this kind of flow. Uniform temperature, mean velocity, and turbulence intensity profiles (Dirichlet boundary condition) were specified for both the main flow and the injected jets at orifice inlets and crossflow entry because experimental data regarding these profiles for non-isothermal impinging *JIC* flows are absent in available literature sources. Wall orifice thickness was 5 mm. Varying the inlet jet turbulence intensity within 3-15% as well as the turbulence intensity at crossflow entry from 0.5 to 5% proved to be insignificant regarding to *JIC* flowfields established for all *J*'s (*C*'s). The mainstream inlet was at *x/D*= -3.0, i.e. positioned much farther than an ultimate upstream penetration of counter flow jet, and the center of the orifices was at *x/D*=0 (*JIP*). Zero-pressure gradient (Neumann boundary condition) was set at the outlet plane which was at *x/D*= 2.5. Temperature was specified on the channel walls from previous tests. The effect of the wall temperature on the core flow of the channel was estimated to be insignificant.

Reynolds number for the crossflow is about 750 while that for jets ~ 17720 for $J^{1/2}$=39.6. The interior diameter of the mainstream duct was *D*=32 mm, and the orifices were *d*=3 mm in diameter (*d/D*=.094). Orifice and channel walls were assumed to be smooth, and no-slip conditions were applied to them.

## 2.4 Simulation background and references

It is of common knowledge that Reynolds-averaged Navier-Stokes based turbulence models were calibrated for statistically stationary flows (mixing layers, jets, wakes, etc.) where all unsteadiness is due to turbulence. In particular, Launder&Spalding [43] demonstrated the range of applicability of the $\kappa$-$\varepsilon$ model by computations of nine substantially different kinds of turbulent flows. Results from $\kappa$-$\varepsilon$ models were also reported by Pope [44] for predicting flows such as an expansion of the submerged circular jet. Previous *JIC* calculations using a steady-state $\kappa$–$\varepsilon$ model (e.g. [2-4]) showed good qualitative agreement with experimental findings, although it is stressed by Sloan et al.[45] that the $\kappa$-$\varepsilon$ model poorly predicted the streamline pattern in reversed flow zones, but gives satisfactory results elsewhere. The *realizable $\kappa$-$\varepsilon$ model* (*RKE*) published by Shih et al.[46], and available in [39] demonstrated good convergence for similar problems. It is especially noteworthy that the *realizable $\kappa$-$\varepsilon$ model* (*RKE*) is capable of resolving the, so-called, the round-jet anomaly; that is, it predicts the spreading rate for axisymmetric jets as well as that for planar jets.

Recently the *realizable $\kappa$-$\varepsilon$ model* (*RKE*) was utilized by Yu et al.[47] for modeling of a *JIC* quenching process using cold pyrolysis for the partial oxidation of methane; and was used by Chen et al.[48] for simulation of an industrial-scale gas quenching process for partial oxidation of natural gas to acetylene. The *realizable $\kappa$-$\varepsilon$ model* (*RKE*) was also used by Kartaev et al.[32] and Kartaev&Holdeman [33]. To the authors' knowledge these publications were the first in which asymmetry was evident and the first use of *RKE* modelling for strongly impinging jets for which the solution is asymmetric and does not converge.

The pressure-based segregated solver that is most suitable for simulating low- and high-velocity incompressible flows was utilized for computing *JIC* flows in this study. The Semi-Implicit Method for Pressure Linked Equations (*SIMPLE*) algorithm of Patankar&Spalding [49] of the pressure-based segregated solver was employed to correct the pressure term interpolated in accord with the standard scheme.

A method based on first-order upwind discretization (*FOU*) was used by McGuirk&Spencer [50&51]. The first-order scheme was chosen because it was reported by Spencer&Adumitroaie [52] that lower-order schemes generally offered the closest agreement to *JIC* experiments. A recent comparison of numerical closure methods appropriate for *JIC*'s was presented by Davoudzadeh et al.[53].

## 2.5 Verification and validation of the numerical model

Formal verification and validation of *RKE* for non-impinging jets for $J^{1/2}$=4.9 (*C*=2.7) &impinging jets for $J^{1/2}$=9.9 (*C*=5.5) without significant backflow are given in [33] and not repeated here. In this, the adequate checks were conducted to justify the choice of the aforementioned grid for the current *CFD* analysis. Also, the credible evidence that the numerical accuracy of *RKE* is suitable for impinging *JIC* flows was obtained in [32].

Holdeman [2] shows 3D plots of experimental data and numerical & empirical model results for various *JIC* configurations. Although the plots in [2] were not assembled to validate the *CFD* results, they seem to show that *CFD* was qualitatively accurate in the 1980's. Holdeman et al.[26] and Holdeman et al.[27] used the trends from the numerical results from steady-state *JIC* calculations in [26] to extend the NASA empirical model for rectangular ducts to annular, reverse-flow, and cylindrical duct configurations [27]. In the 1990's *CFD* results were much more quantitative. Holdeman et al. [3] & Holdeman et al. [4] show both experimental and numerical results to identify features of *RQL* mixers in cylindrical [3] and rectangular [4] ducts. Note that the figures in [2-4] are much better in color in the original papers that are referenced in [2-4].

*URANS* cannot be formally verified as it is an unsteady solver. However, *URANS* and *RANS* results are virtually identical in the *JIC* steady-state case for $J^{1/2}$≤9.9 (*C*≤5.5) [33].

It is known that *URANS* simulations allow one to study turbulent flows with very large-scale unsteadiness at low enough frequency that it's time scale is lower than slowest time scale represented by large scale turbulent eddies, effects of these eddies are modelled as the Reynolds stresses [40]. In addition, eddy-viscosity models may display substantial deficiencies in massively separated flows. Nonetheless, it seems reasonable to claim that *URANS* is capable to resolve properly slow large-scale ("deterministic") *JIC* flow motions in the case of strongly interacting jets, for instance, observed experimentally in [34].

For *URANS* simulation of unsteady *JIC* flows the pressure-based segregated approach implies the fully implicit temporal scheme of $2^{nd}$ order accuracy that is unconditionally stable with respect to time step size. For computations time step size of $10^{-3}$ s was set, with total number of time steps of 5000 in order to detect self-induced low-frequency *JIC* flow motions (>0.2 Hz). The same temporal scheme with time step of $10^{-3}$ s was also applied for *URANS* simulations of air swirl flows in fuel injectors [54]. It should be notified though that for such flows with very low-frequency (~ 1Hz) unsteadiness the impinging *JIC* flows obtained using the fully implicit temporal schemes with $2^{nd}$ and $1^{st}$ orders of numerical accuracy turned out to be almost identical.

*RKE* results for strongly impinging jets cannot be validated because there are no asymmetric data for strongly impinging jets. Hence, *TA* was added to verify the integrity of our unsteady *CFD* results. Also we added references to related experimental evidence of unsteadiness, such as Thong et al. [34], Spencer [55] and Hollis [56].

Using *RKE* and *URANS* for strongly impinging *JIC* is an extrapolation for the model as the conditions are outside the range of statistically steady flows for which the model can be verified and validated. Any model invoked for strongly impinging jets would be an extrapolation because experimental data regarding unsteady *JIC* flows are scarce (e.g. [34], [55], [56]).

## 3 Results and Discussion

The numerical results obtained for a counter flow *JIC* mixer in a cylindrical duct are consistent with experimental and numerical results motivated by mixing in gas turbine combustors [2-4]. In fact, the region that appeared to be unsteady in [32] and [33] had *C* values greater than that designated as "overpenetrating" in previous studies. From [2], $C=\pi(2J)^{1/2}/n$ for a cylindrical duct, where *J* is the momentum-flux ratio, and *n* is the number of orifices. Note that for 8 jets $(S/H)_{can}=.555$, thus $C=0.555(J^{1/2})$. Also, since the orifice to duct diameter ratio, and the jet-to-mainstream density ratio ($d/D$=3/32; *DR*=5) were also constant in this study, $MR=0.158(J^{1/2})$. Thus, for the flow and geometry conditions in this study, it is "dealer's choice" as to whether *J, C,* or *MR* is the primary parameter. (Based on previous results, the authors think that *C* is the most important parameter for identifying the expected jet penetration.)

## 3.1 Case of $J^{1/2}$=39.6

Fig.2 shows a vertical centerplane for a configuration with strongly impinging jets which result in a counter-flow for jets injected from the perimeter of the cylindrical duct with the radial velocity $V_j$. Fig.2 is similar to Fig.1b in [32], but with the following significant differences:

1) Fig.2 is the vertical centerplane through jets 3&7, whereas Fig.1b shown in [32] is the horizontal centerplane through jets 1&5. Because the horizontal centerplane was a plane of symmetry in [32], jets 3&7 would be the same.
2) As recommended in [33], the calculations in Fig.2 used time-dependent *URANS*, whereas Fig.1b in [32] used steady-state *RANS*.
3) As also recommended in [33] (critical for strongly impinging jets), the calculation domain in Fig.2 was the complete cylindrical duct whereas that in Fig.1b in [32] was half the cylindrical duct.
4) Fig.2 includes the temperature distribution whereas Fig.1b in [32] did not.

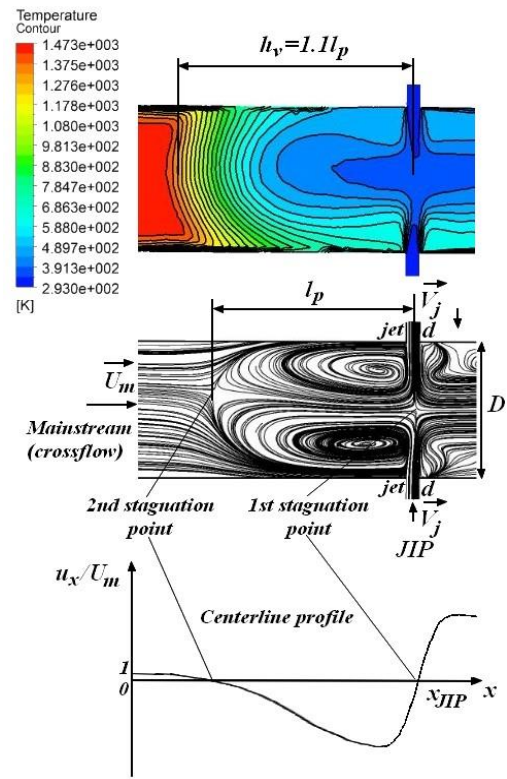

**Fig. 2** Radial-axial plane schematics of the *JIC* configurations of the counter flow jet formed in the mainstream by impingement of eight *JIC*'s confined by cylindrical duct walls. A profile of the dimensionless velocity on the duct centerline is also shown. *URANS* results for $J^{1/2}=39.6$ in 360° computational domain at the moment $t=0.730$ sec

$X=0$ at the orifice center so distances upstream and downstream are negative and positive, respectively from the center of the jets (*JIP*). Fig.2 also shows the flow velocity $u_x$ on the duct centreline normalized by the velocity $U_m$ of the mainstream inlet. The positive direction of the *x* axis coincides with the inlet velocity $U_m$ direction, thus the flow velocity $u_x$ on the duct axis is positive initially in the approach (mainstream) flow and is negative in the recirculation flow zone (*RFZ*) upstream of *JIP*.

For impinging *JIC*'s, two points of flow stagnation form on the channel centreline. The 1st stagnation point (near *JIP*) divides the upstream and downstream flow from the impinging jets, and is located downstream of the center of the jet orifices at low jet-to-mainstream momentum-flux ratios. However, for strongly impinging jets, this stagnation point is almost coincident with the location of the jet injection plane (*JIP*). A second stagnation point on the duct centreline occurs at the maximum upstream penetration of a counter flow jet. The penetration depth $l_p$ of the counter-flow jet is the axial distance between the farthest upstream flow stagnation point and *JIP*.

For the case of mixing of non-isothermal flows the dimensionless upstream axial penetration depth $h_v/D$ of cold counter-flow jet into the high-temperature mainflow was first introduced in [28,29]. It was measured on the base of the temperature decrease by a certain preset value which was taken to be 100 K. The calculations gave $h_v \cong 1.1 \ast l_p$.

As seen from Fig.2, the counter-flow jet first accelerates and then is decelerated by the mainstream flow. The flow on the centerline opposes the mainstream flow. A toroidal recirculating flow zone (*RFZ*) forms between the stagnation points and is characterized by the 'eyes' of zero velocity. Note that in Fig.2 the counter-flow jet is not perfectly symmetric; i.e. jets 3&7 are not the same. This snapshot is at the time for the farthest upstream penetration of the counter-flow jet and is prior to the onset of 'flapping' (a term from [34]). The observation of asymmetry in [32&33] and unsteadiness reported from the experiments in [34] for high *J* motivated the current study.

## 3.2 *URANS* solution for $J^{1/2}$=39.6 at the instant just before onset of *JIC* 'flapping'

Figs. 3&4 show *URANS* solution for $J^{1/2}$=39.6 in the 360° computational domain at *t*=0.730 seconds corresponding to the farthest upstream centerline penetration of the counter-flow jet and is before the onset of 'flapping' of *RFZ*. Fig.3 shows all centerplane and midplane *r-x* distributions. Note that Fig.3e is the condition shown in Fig.2. Fig.4 shows the corresponding *r-θ* plots at several locations between the impingement and farthest upstream stagnation points.

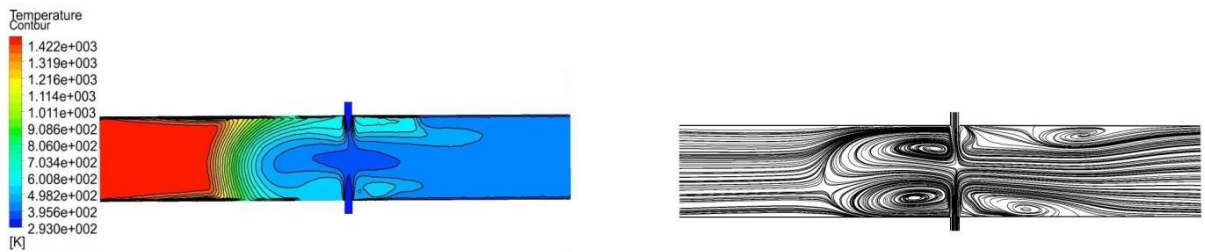

a) centerplane - jets 1 & 5

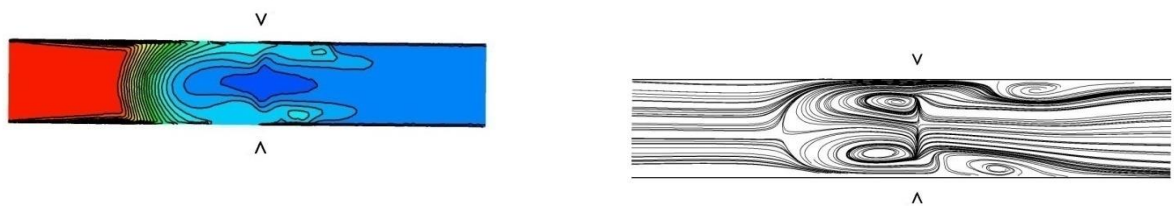

b) midplane 1.5 & 5.5

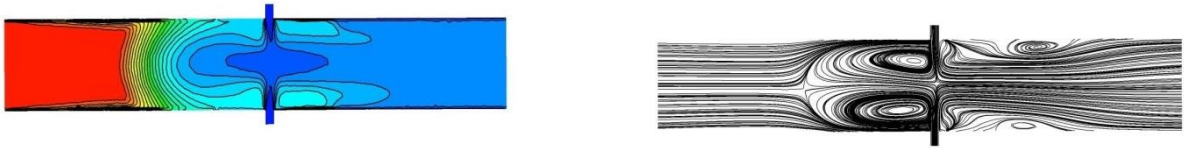

c) centerplane - jets 2 & 6

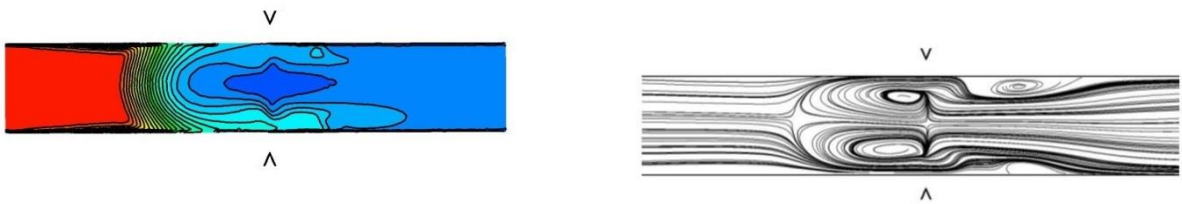

d) midplane 2.5 & 6.5

**Fig. 3 (Part 1 of 2)** Unsteady *URANS* centerplane (*r-x*) plots for 360° calculation domain; *n*=8, *DR*=5, $J^{1/2}$ =39.6 (*C*=22.0, *MR*=6.26), *t*=0.730 sec. Checkmarks in the midplane plots indicate the axial location of *JIP*

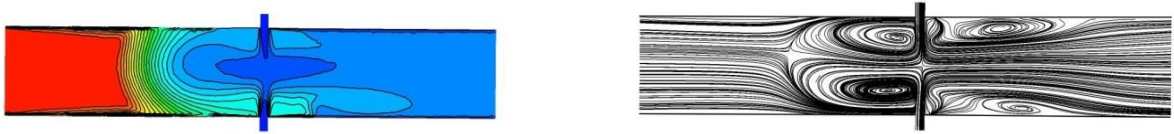

e) centerplane - jets 3 & 7

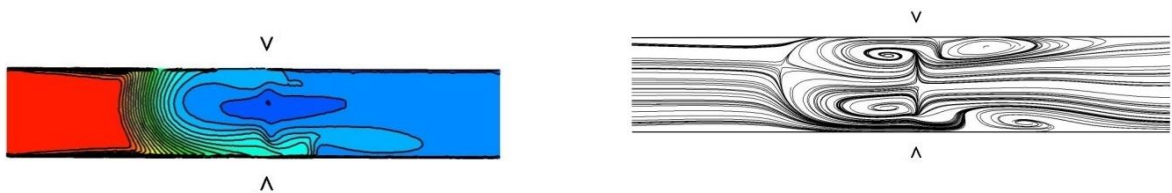

f) midplane 3.5 & 7.5

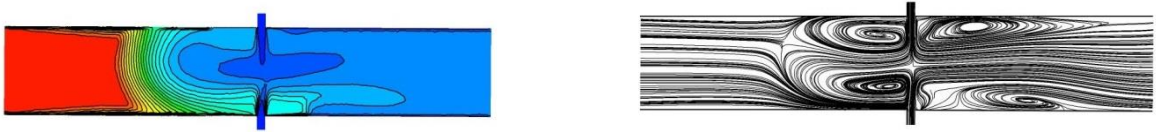

g) centerplane – jets 4 & 8

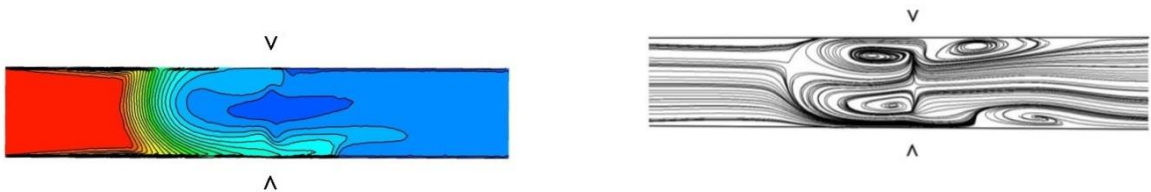

h) midplane 4.5 & 8.5

**Fig. 3 (Part 2 of 2)** Unsteady *URANS* centerplane (*r-x*) plots for 360° calculation domain; *n*=8, *DR*=5, $J^{1/2}$=39.6 (*C*=22.0, *MR*=6.26), *t*=0.730 sec. Checkmarks in the midplane plots indicate the axial location of *JIP*

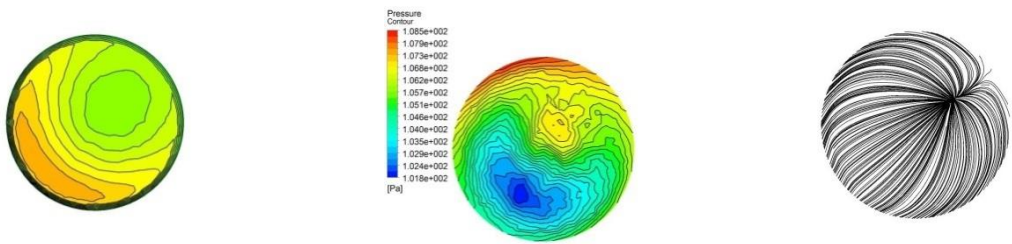

a) *x/D*= - 1.4

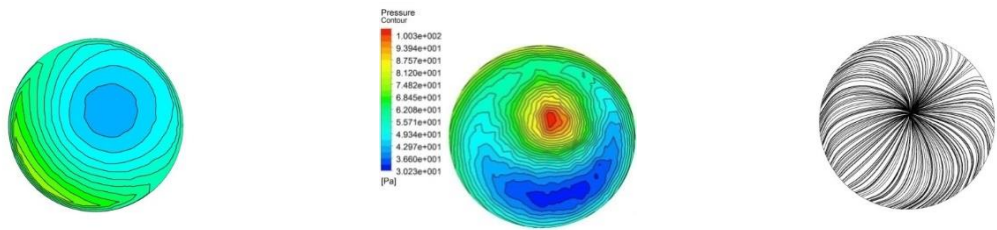

b) *x/D*= - 1.0

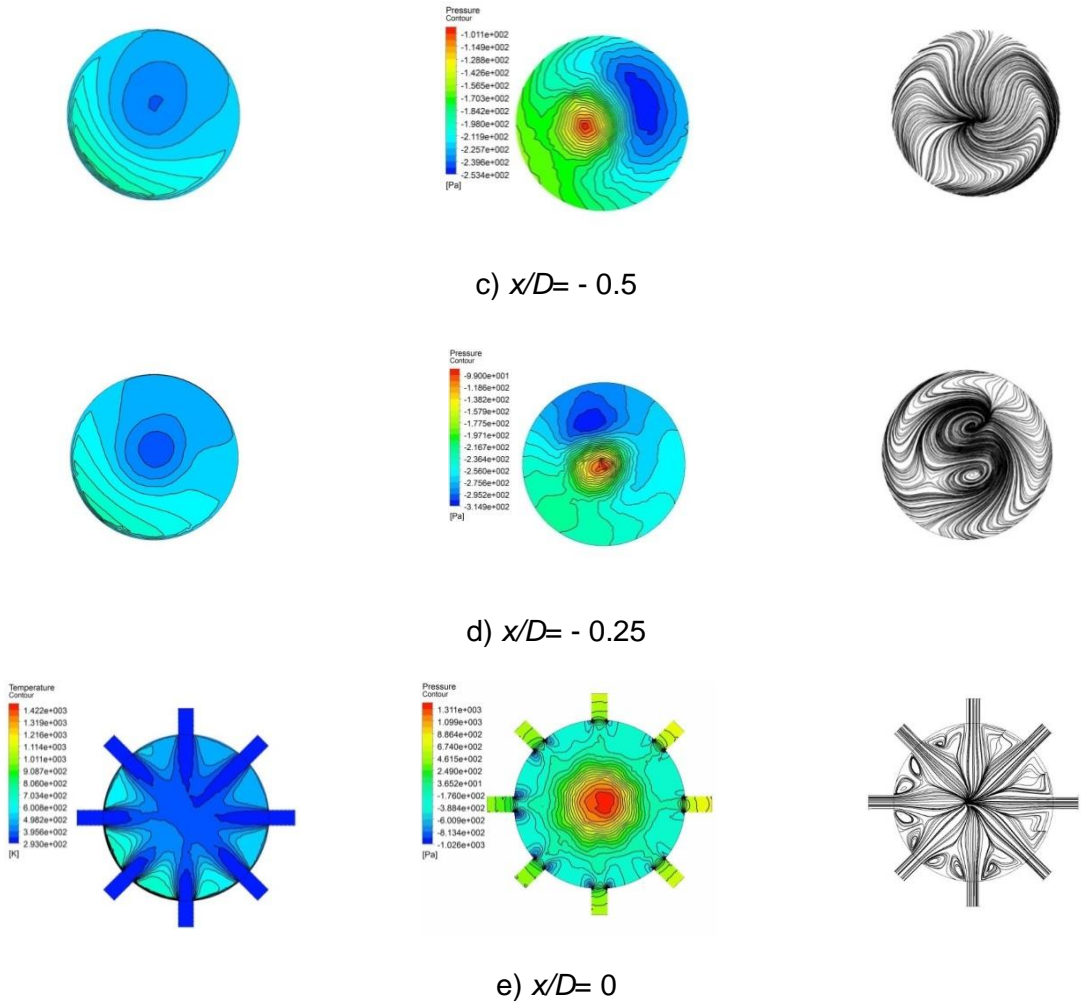

c) *x/D*= - 0.5

d) *x/D*= - 0.25

e) *x/D*= 0

**Fig. 4** Unsteady *URANS* radial-transverse (*r-θ*) plots for 360° calculation domain, *DR*=5, $J^{1/2}$=39.6 (*C*=22.0, *MR*=6.26), *t*=0.730 sec. The first column is temperature; the second column is gauge pressure; the third column is instant (pseudo-) streamlines

It is seen from the streamline plots in Fig.3 that in addition to the above mentioned upstream recirculating flow zone (upstream *RFZ*) there is a recirculating flow zone downstream of *JIP* (*downstream RFZ*). It is witnessed by the presence of zero-velocity 'eyes' of both *RFZ*s simultaneously in all centerplanes (jets' planes) and midplanes. Both *RFZ*s appear to have already an asymmetrical (deformed) torus-like shape at this instant before the onset of *JIC* 'flapping'.

Also, it may be seen from Fig.4 that *JIC* flow at *JIP* (*x/D*=0) is still almost symmetrical whilst the upstream *JIC* flow field nearby 2nd stagnation point (*x/D*=-1.4) already shows a prominent asymmetry.

## 3.3 *URANS* solution for $J^{1/2}$=39.6 after establishment of *JIC* 'flapping'

Figs.5&6 show a *URANS* solution at *t*=3 seconds for $J^{1/2}$=39.6 (*C*=22.0; *MR*=6.26) using the 360° (full cylindrical duct) computational domain.

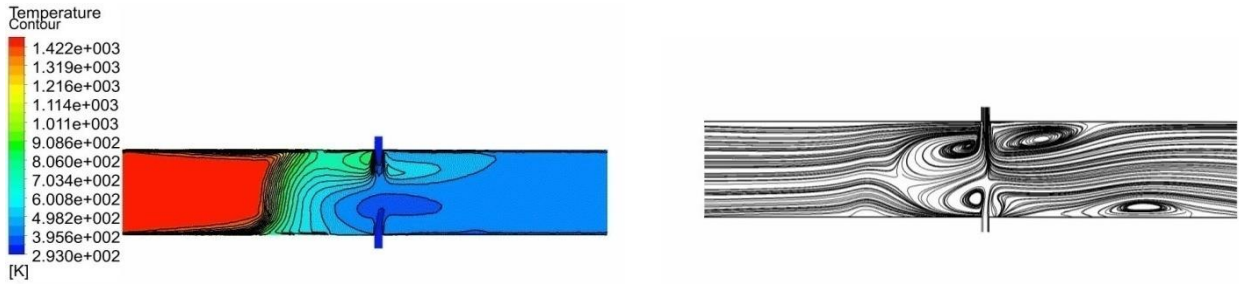

a) centerplane – jets 1 & 5

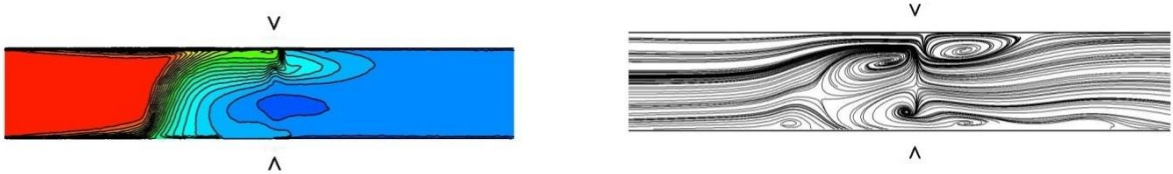

b) midplane 1.5 & 5.5

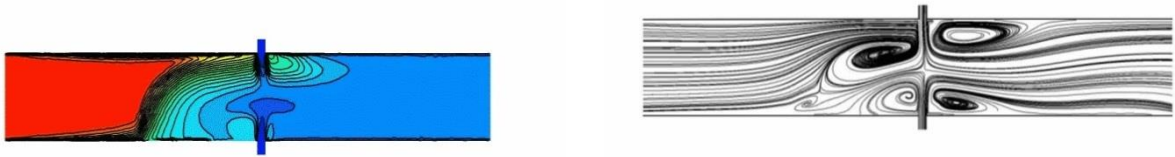

c) centerplane – jets 2 & 6

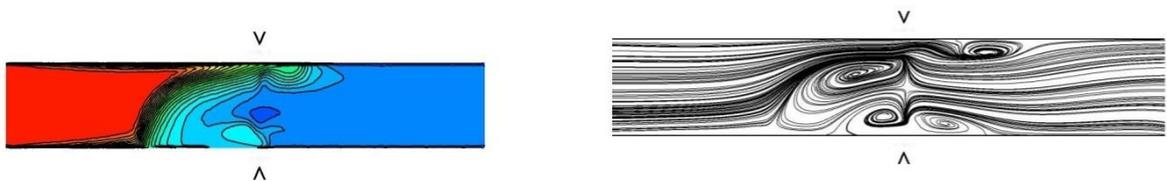

d) midplane 2.5 & 6.5

**Fig. 5 (Part 1 of 2)** *URANS* *r-x* plots for 360° calculation domain; *n*=8, *DR*=5, $J^{1/2}$=39.6 (*C*=22.0, *MR*=6.26), *t*=3.000 sec. Checkmarks in the midplane plots indicate the axial location of *JIP*

The difference between Figs.3&4 and Figs.5&6 is time. Figs.5&6 at *t*=3 sec are after flapping is established and regular. As in Fig.3, Fig.5 shows all the jet and midplane pairs from jets1&5 to midplane 4.5&8.5. However, this is really only half the cylindrical duct; the other half, from jets 5&1 to midplane 8.5&4.5, are not shown as they are upside-down versions of those in Figs.5a-h.

Because the *URANS* solutions in the 360° computational domain show significant asymmetry, steady-state *RANS* computations would not provide an acceptable representation of the flow field.

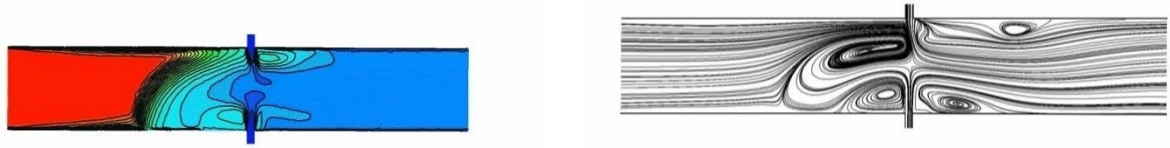

e) centerplane = jets 3 & 7

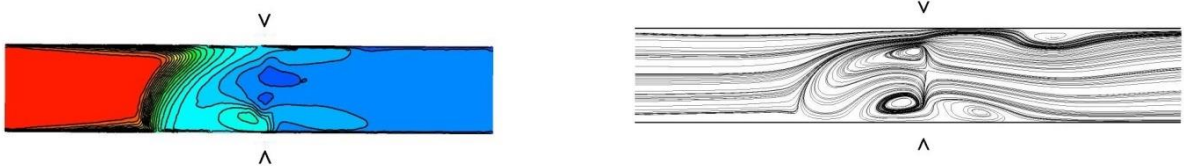

f) midplane 3.5 & 7.5

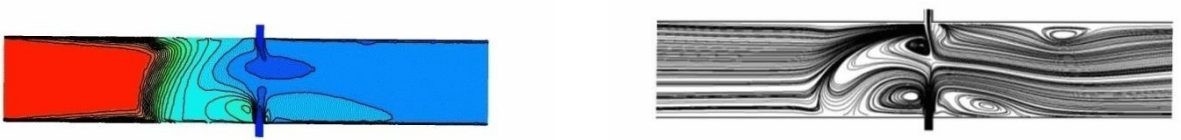

g) centerplane – jets 4 & 8

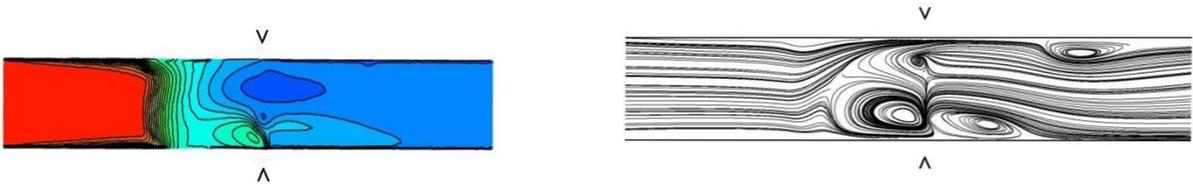

h) midplane 4.5 & 8.5

**Fig. 5 (Part 2 of 2)** *URANS* centerplane (*r-x*) plots for 360° calculation domain; *n*=8, *DR*=5, $J^{1/2}$=39.6 (*C*=22.0, *MR*=6.26), *t*=3.000 sec. Checkmarks in the midplane plots indicate the axial location of *JIP*

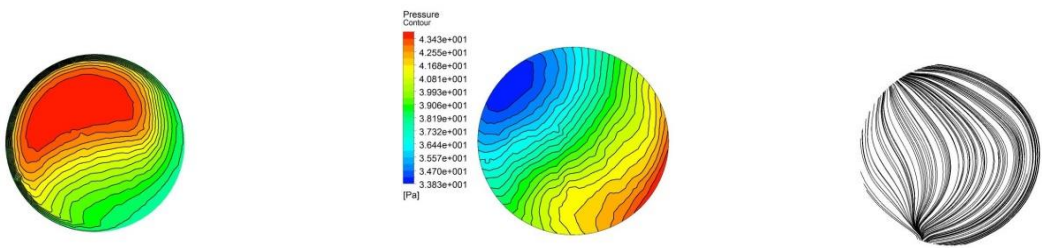

a) *x/D*= - 1.3

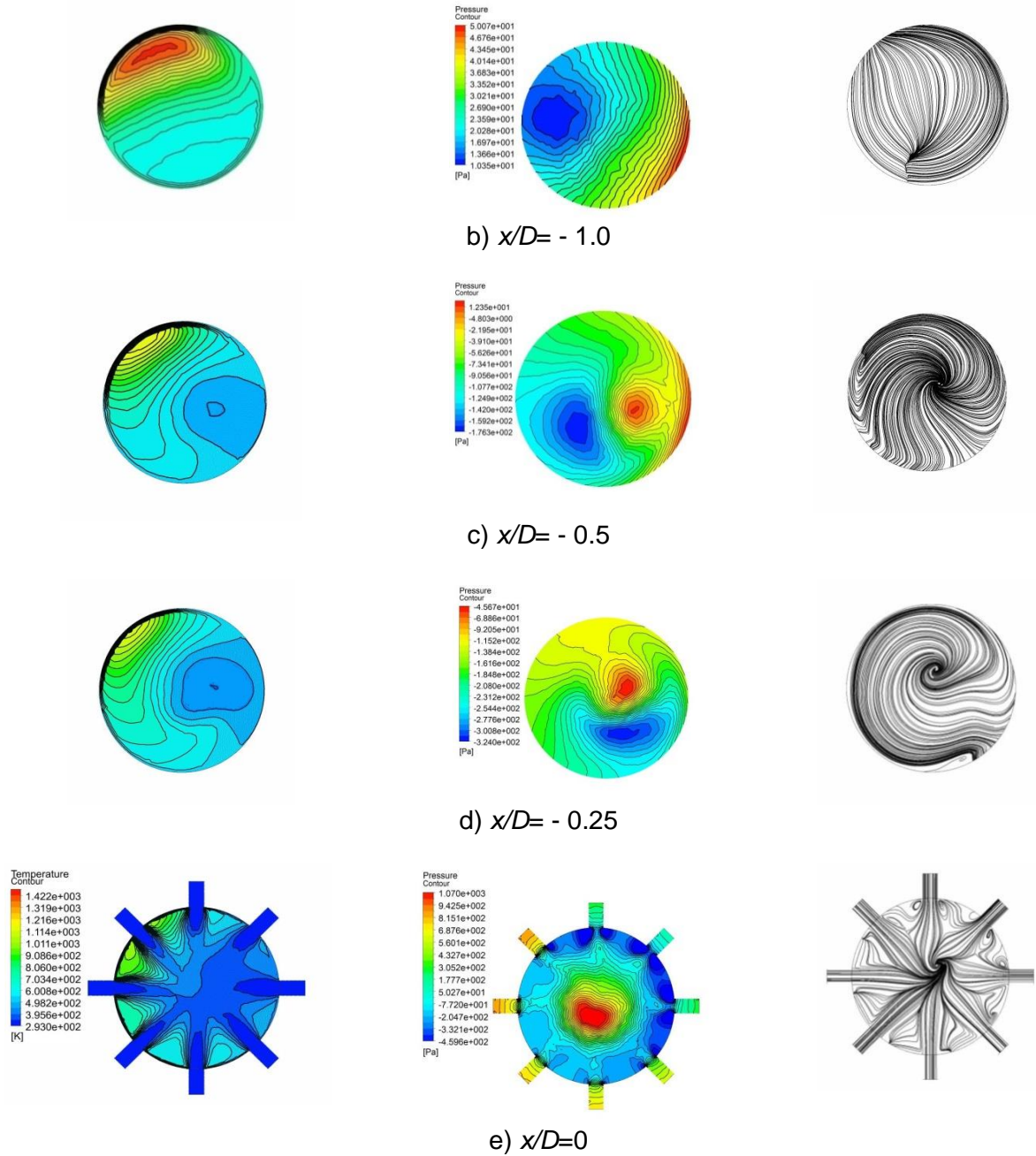

b) *x/D*= - 1.0

c) *x/D*= - 0.5

d) *x/D*= - 0.25

e) *x/D*=0

**Fig. 6** *URANS* radial-transverse (*r-θ*) plots for 360° calculation domain; *n*=8, *DR*=5, $J^{1/2}$=39.6 (*C*=22.0, *MR*=6.26), *t*=3.000 sec. The first column is temperature; the second column is gauge pressure; the third column is instant (pseudo-) streamlines

Likewise it may be noticed that both upstream&downstream *RFZ*s are formed on the streamline plots shown in Fig.5. The difference between streamlines drawn in Fig.3 and those in Fig.5 is that after establishing regular *JIC* 'flapping' the torus-like shape of both upstream&downstream *RFZ*s became even more asymmetrical (deformed).

In contrast to *JIC* flowfield depicted in Fig.4, it is evident from Fig.6 that *JIC* flows at *JIP* (*x/D*=0) and nearby 2nd stagnation point (*x/D*=-1.4) are visibly asymmetrical.

A possible reason for the significant asymmetry in Figs. 5&6 is: as *J* increases, the pressure near the duct centerline at *JIP* increases. However, in a cylindrical duct the mixing region associated with each jet decreases in width as the jet centerline is approached from the perimeter. Thus, the jets cannot spread laterally to relieve the increasing pressure near 1st stagnation point at *JIP*. Eventually, for strongly impinging jets, the symmetric, steady flow becomes unsteady. This effect can also be explained by the local growth of pressure gradient for high *J*'s in the vicinity of the second stagnation point at the maximum upstream penetration of a counter flow jet. Local peaks of pressure at both stagnation points are seen on the duct centerline profiles of gauge pressure before onset of *JIC* 'flapping' (term from [34]) in Fig.15 of [32].

A plausible explanation of unsteadiness driving asymmetry may be found from a comparative analysis of *JIC* flowfileds shown in Figs.4&6. From Fig.4 it may be inferred that prior to onset of *JIC* 'flapping' an asymmetry initially occurs at the 2nd stagnation point while flowfield at *JIP* is rather symmetrical. At the moment after establishment of *JIC* 'flapping' shown in Fig.6 flow asymmetry is clear at the 2nd stagnation point and at *JIP*. It may be assumed that an initial flow asymmetry propagates from the 2nd stagnation point to *JIP* (1st stagnation point) and farther downstream.

From the point of *JIC* mixing effectiveness it may be assumed that most of high-temperature mainflow is mixed with cold jets within *RFZ* upstream of *JIP*. An unmixed part of mainflow penetrated between jets near duct walls is envolved to *RFZ* downstream of *JIP*. This flow configuration actually provides intensive &localized *JIC* mixing.

Because the recirculation regions of the opposed jets are not the same, the counter-flow jet has deflected off the duct centerline with the onset of 'flapping' (term from [34]).

### 3.4 Cycling
### 3.4.1 Cyclic solutions for $J^{1/2}$=39.6 (*C*=22.0; *MR*=6.25)

In Figs.7&8, temperature, pressure, and streamline plots from 360° *URANS* calculations are shown in radial-azimuthal (*r-θ*) planes for $J^{1/2}$=39.6 (*C*=22.0; *MR*=6.25) at several time slices. Note that the *r-θ* temperature plots at *x/D*= - 0.5 and 0 at *t*=3 sec in Fig.7 are also shown in Fig.6. Fig.7 shows that the variation with time is cyclic at a frequency of approximately 1.8-2.0 Hz (estimated from the temperature plots at *x/D*= - 0.5 in Fig.7). A cyclic motion with the same frequency was observed for $J^{1/2}$= 13.2 and 20.7 (*C*=7.3&11.5; *MR*=2.08&3.27), so it appears that the frequency is not a function of the momentum-flux ratio *J* or this dependence is very weak.

| *Time slice, sec* | *x/D = -0.5* | *x/D=0* |
|---|---|---|
| 2.355 | 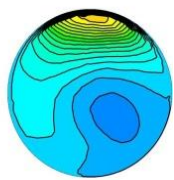 | 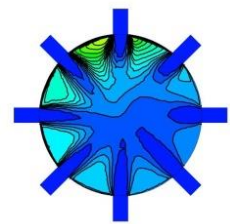 |

*2.450*

*2.500*

*2.640*

*2.734*

*2.870*

*3.000*

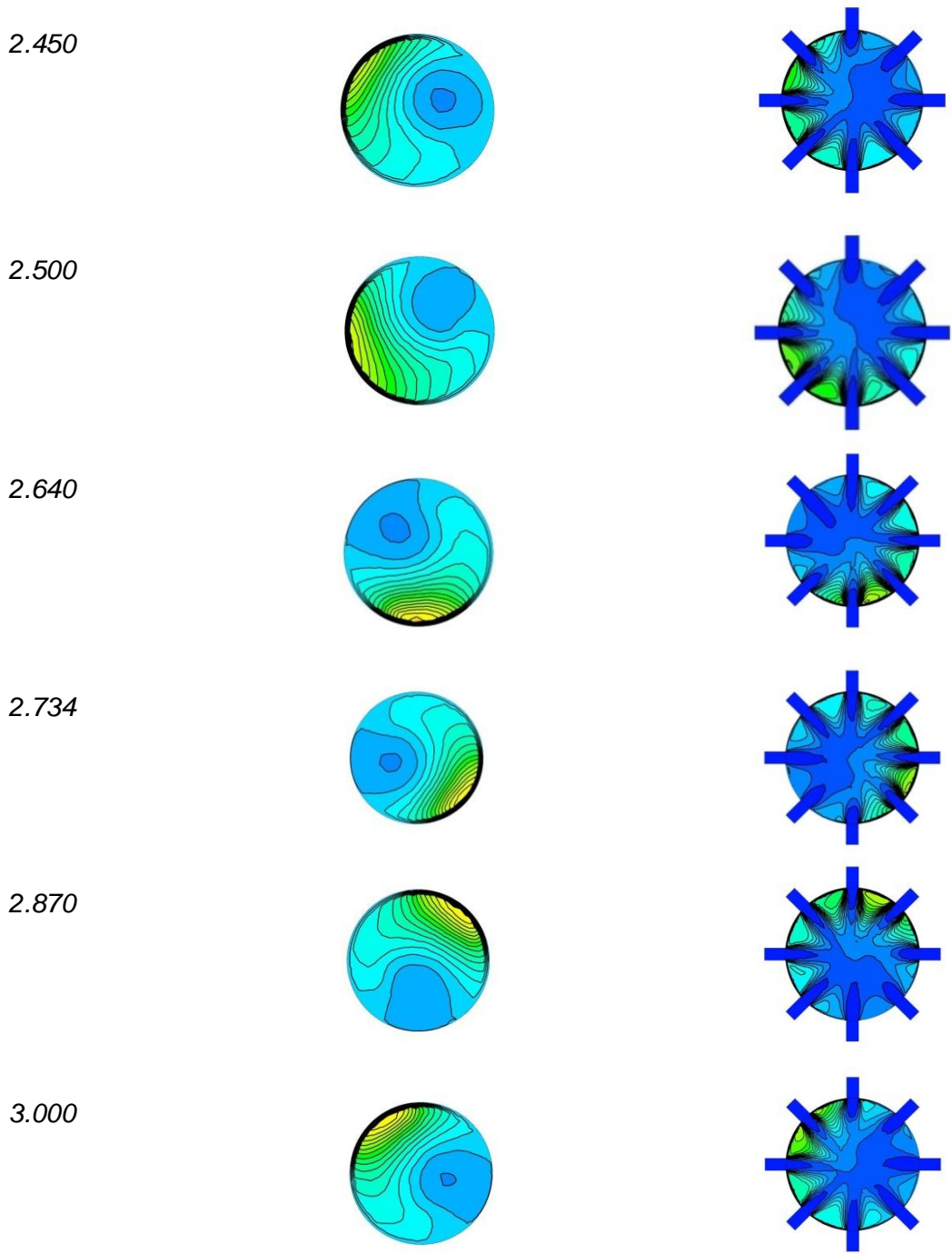

**Fig. 7** *URANS* radial-transverse ($r$-$\theta$) temperature plots for 360° calculation domain; *n*=8, *DR*=5, $J^{1/2}$=39.6 (*C*=22.0), *MR*=6.25 for different time slices

The cycling shown in Fig.7 is confirmed by the pressure and streamline plots at *x/D* = - 0.5 in Fig.8, where the highest pressures coincide with the lowest temperatures. Note that the core of the counter-flow jet is no longer in the center of the cylindrical duct. The cycling appears to be counter-clockwise but there is no obvious reason for a preferred direction of cycling in this case as there would be in a swirling flow or with *JIC*'s from slanted slots. The authors think that the direction of cycling depends rather on fluctuations that ultimately force the developed counter-flow jet off the duct centerline.

It should be emphasized that pressure rise nearby leading edges of orifices does not result in substantial distortions across the jets inlets.

The results for strongly impinging jets revealed large-scale instability of the *JIC* flow field at high *J*'s (*C*'s) which should be accounted in the further analyses of thermal and mechanical loads of components downstream of the mixer, such as the turbine in *GTE*'s.

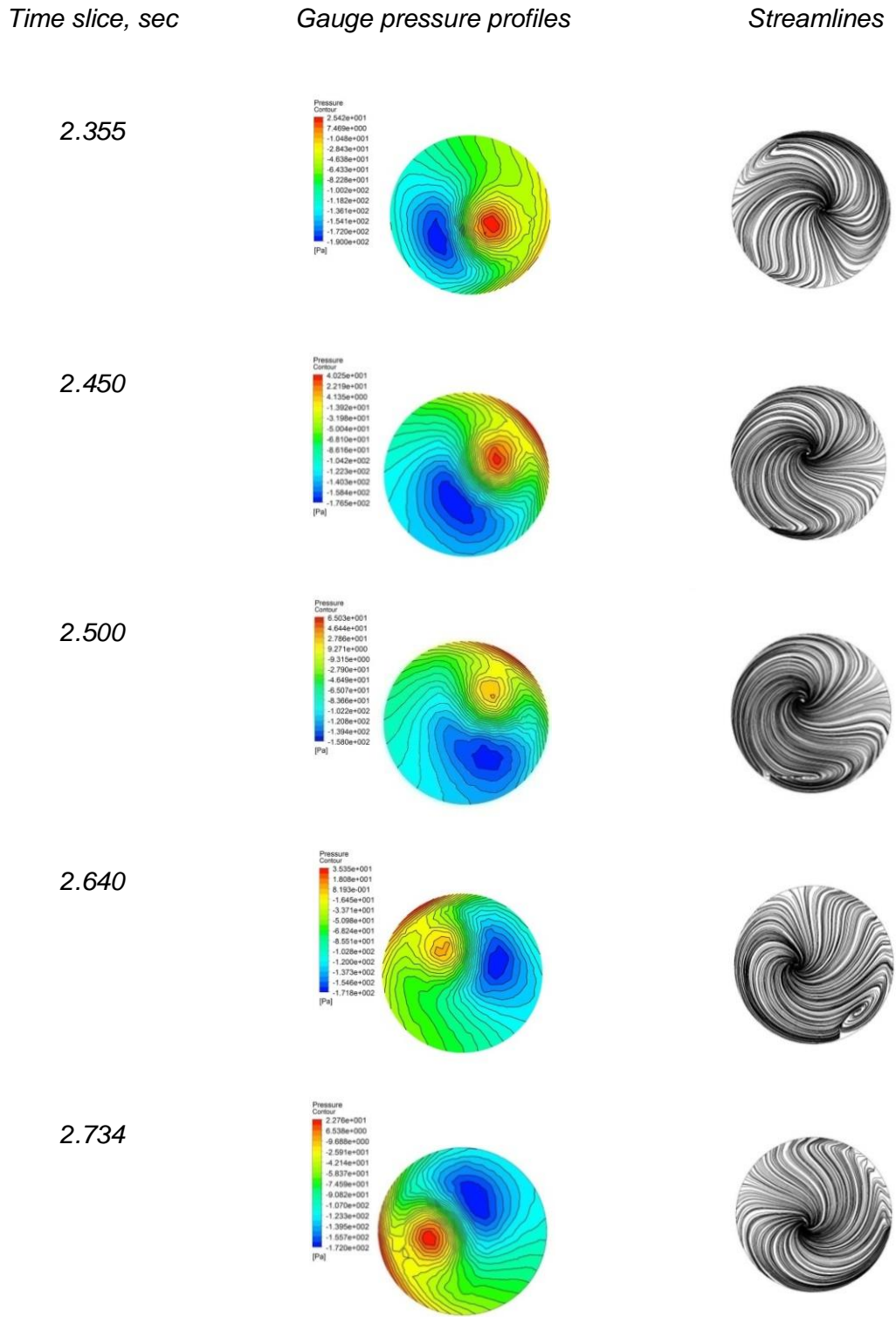

| Time slice, sec | Gauge pressure profiles | Streamlines |
|---|---|---|
| *2.355* | | |
| *2.450* | | |
| *2.500* | | |
| *2.640* | | |
| *2.734* | | |

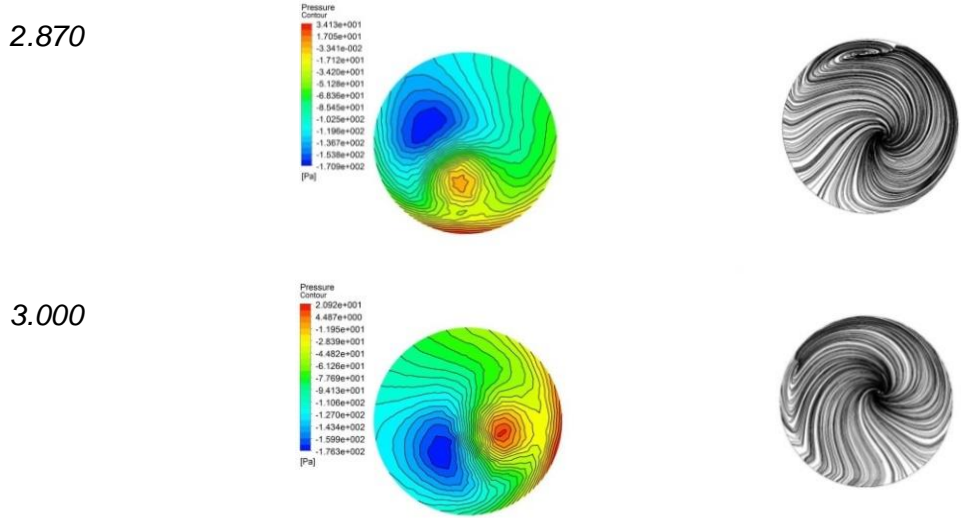

**Fig. 8** *URANS* radial-transverse ($r$-$\theta$) local gauge pressure and streamlines plots at $x/D= -0.5$ for 360° calculation domain; $n=8$, $DR=5$, $J^{1/2}=39.6$ ($C=22.0$), $MR=6.25$ for different time slices

The results shown in Figs.7&8 indicate that the impingement point of mainflow and the counter flow jet (2$^{nd}$ stagnation point) appears not only wanders axially backwards and forwards but also moves radially and azimutally. Thus, for the case of multiple, impinging jets and non-swirling crossflow a complicated *JIC* cycling (3D unsteadiness) occurs because of an inherent instability of this flow configuration.

### 3.4.2 On estimates of cycling frequency from *URANS* calculations to the physically observed phenomena

Although experimental information for 'flapping' of impinging *JIC* flow in non-isothermal flows is not available, the qualitative analysis of accuracy of the obtained numerical results may be carried out based on data of *JIC* water tests by Spencer [55] and Hollis [56].

In both works the authors studied experimentally transient flow structures that occurred as a result of impingement of multiple jets into the mainstream flow of the cylindrical duct. These flow structures are hardly identifiable as they are invisible in the time average. In particular, the transient bi-modal 'flapping' behavior of upstream recirculation flow zone (*RFZ*) was studied in [55] by means of flow visualization. Such a behavior is visually expressed as periodic considerable change of size of *RFZ*. To illustrate this, the photographs of flow visualization taken approximately 3 sec apart were shown in Plate 3.1 of Spencer [55]. Similarly, a substantial change of *RFZ* size can be observed for example from comparison of predicted *JIC* streamlines shown in Fig.3a ($t=0.73$ sec) and in Fig.5a ($t=3.00$ sec).

Hollis [56] was tracing a bi-modal behavior of velocity vector flowfields at the selected duct centerplane based on high-quality LaVision *PIV* system. Experimental data were acquired with time step of 0.0667 sec in that plane. From Fig.7.42 in Hollis [56] it can be seen that counter flow jet (formed by impingement of multiple jets) alters its direction from one duct wall to the opposite one in this plane after 4 time steps, i.e. during 0.267 sec. It can be roughly estimated as the half-period of cyclic motion of counter flow jet (as well as that of *RFZ* 'flapping') since it could be considered as 'a projection' of 3D cyclic motion of *RFZ* on this centerplane. Therefore, the period of cyclic motion of *RFZ* appears to be close to 0.5 sec. In addition, in the Fig.7.42 it is seen a formation of near-wall boundary layer vortex separately from *RFZ* 'sticked' to a wall. In our numerical study the low-frequency motion of *RFZ* was revealed from Figs.7&8. The *URANS*-predicted near-wall vortex separately from *RFZ* is also clearly seen in streamlines depicted in Fig.5a-c. Remarkably, time average flow topology (streamlines and *RFZ*) presented in Figs.7.18&7.19 of [56] resembles the *JIC* flowfield pattern shown in Fig.2 before the onset of *RFZ* flapping. Hollis' frequency estimates not only agree with our *URANS* predicted ones, but they seem to confirm our (unpublished) conclusion that frequency is not a function of *J*.

Based on the above mentioned analogies it appears that the *RKE* model is capable of predicting the large-scale 'deterministic' (non-turbulent) phenomena in the case of strongly impinging *JIC* such as an occurrence of flow asymmetry & instability, low-frequency *RFZ* 'flapping' and formation of near-wall vortices separately from *RFZ* upstream of *JIP*.

For the conditions of these experiments the *LES* studies reporting unsteady nature of *JIC* flow fields were first carried out by Spencer&Adumitroaie [52] and later by Clayton&Jones [57]. Also, for the same *JIC* flow configuration the high-resolution *LES* study by Saini, Xia&Page [58] has been recently carried out. Due to imposing refined grids the latter simulation enabled to resolve such periodic flow features as a motion of jet shear layer vortices. *LES* predictions of [58] indicate *JIC* axial velocity oscillations of frequency of ~1Hz upstream of *JIP*. Time animation of such *JIC* flow oscillations may be extracted from [59]. In aforementioned *LES* studies this slow *JIC* 'flapping' was observed only in the jet centerplane without unveiling 3D complex motion of such impinging *JIC*s. Our low-cost *URANS* calculations seemingly allowed one to follow some of 3D large-scale *JIC* flow features.

## 3.5 Topological analysis for $J^{1/2}$=39.6
### 3.5.1 Basics of *TA*

In applied mathematics, topological analysis (*TA*) is an approach to the analysis of datasets using techniques from topology. Although the current dataset is a study in fluid mechanics, the topology conditions are not based upon fluid or solid mechanics; just a vector field overlaid on a surface. Foss [36] and Foss et al.[37] establish the procedures to evaluate the integrity of a measured or computed vector field that is overlaid on a designated surface. The essential content from these two references is summarized here to form the basis for an evaluation of the computed *JIC* results.

Any surface can be formed from a sphere by opening holes through the surface and adding handles to it. A constraint on the designation of a hole is that the vector field at the perimeter of the hole must be directed onto or away from the surface. Once formed, the Euler characteristic ($\chi$) for the surface is equal to

$$\chi = 2(\text{for the sphere}) - \sum \text{holes} - 2\sum \text{handles} \qquad (1)$$

Since this value is known a' priori, it is further identified as $\chi_a$.

The computed or experimental vector field may exhibit singular (critical) points which will have the designation of a node (*N*) or a saddle (*S*). Foss [36] provides the technique to identify each of these using a unit vector which translates clockwise around the candidate point with its base touching that small circle and the unit vector aligned with the vector field at each point on the circle. If the unit vector rotates clockwise once in that transit, the circled point is a node. If it rotates once counter-clockwise a saddle has been circled. If there is no rotation, the candidate is not a singular point. A nodal point has an index of +1. A saddle index is -1. The sum of the indices is equal to the Euler characteristic for the designated surface. That is,

$$\chi_e = \sum N - \sum S \qquad (2)$$

where the subscript $e$ is used to show its relationship to an experimental or computational observation.

The surfaces of present interest have the character of a "collapsed (deflated) sphere". That is, a planar surface whose perimeter is aligned with the subject vector field (i.e., a streamline in the present case). Since interior singular points (designated as "full nodes and saddles") would be present on both sides of the collapsed sphere, they count twice. A half-node ($N'$) or half-saddle ($S'$) on the perimeter counts once. Equation 2 is therefore written as

$$\chi_e = 2\left(\sum N\right) + \sum N' - 2\left(\sum S\right) - \sum S' \qquad (3)$$

for a 'collapsed' sphere.

A special case addition to (3) is required if the perimeter results in an obtuse angle; see Foss et al.[37] for its careful explanation. Specifically, $\chi_e$ is reduced by 1 for each obtuse angle. The integrity of the experimental vector field is then evaluated by the condition: does $\chi_e = \chi_a$?

### 3.5.2 *TA* results

The 'collapsed sphere' (similar to a deflated balloon) for the *r-x* centerplane, *r-x* midplane, and the *r-θ* plane at *x/D* = -0.5 are shown in Fig.9a-c for *t*=0.73 sec.

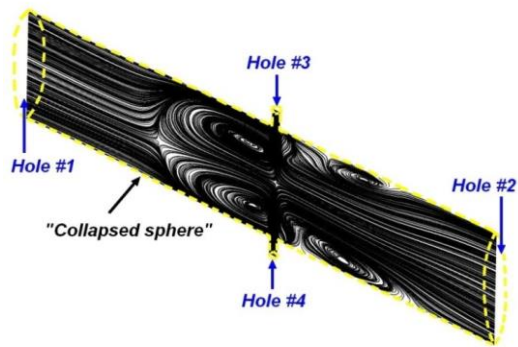

a) 4-hole 'collapsed' sphere (centerplane jets 3-7)
 $\chi_a$ = 2 - Σholes - 2Σhandles = 2-4-2*0 = -2

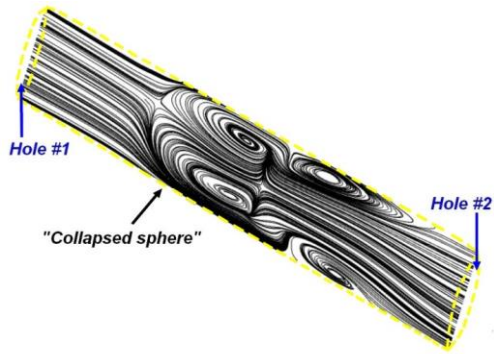

b) 2-hole 'collapsed' sphere (midplane 3.5-7.5)
$\chi_a$ = 2 - $\Sigma$holes - 2$\Sigma$handles = 2-2-2*0 = 0

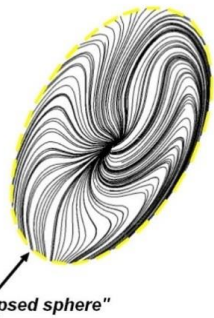

c) 'Collapsed' sphere with no holes (r-$\theta$ plane at x/D= - 0.5)
$\chi_a$ = 2 - $\Sigma$holes - 2$\Sigma$handles = 2-0-2*0 = 2

**Fig. 9** Isometric views of the three shapes of the 'collapsed' spheres for *URANS* streamline plots for 360° calculation domain; *n*=8, *DR*=5, $J^{1/2}$ =39.6 (*C*=22.0, *MR*=6.26), *t*=0.730 sec

Note that the 'collapsed sphere' doesn't change with computational time. Because there are no obtuse corners in $\chi_e$, the zeros for them are not reported in this *TA*.
  Topology analysis of the streamline plots is presented in Figs.10-13. Data plots are also shown in Figs.3-6.

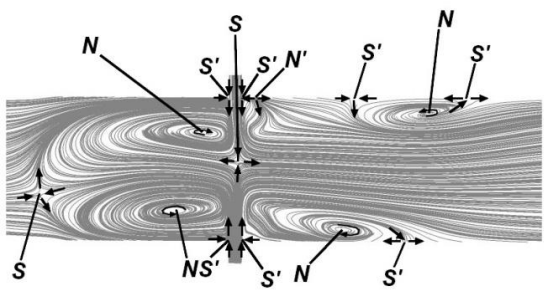

$\chi_e = 2\Sigma N + \Sigma N' - 2\Sigma S - \Sigma S' =$
     $= 2*4 + 1 - 2*2 - 7 = -2$

a) centerplane - jets 1&5

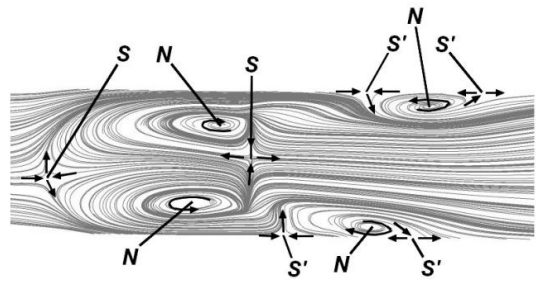

$\chi_e = 2\Sigma N + \Sigma N' - 2\Sigma S - \Sigma S' =$
     $= 2*4 + 0 - 2*2 - 4 = 0$

b) midplane 1.5&5.5

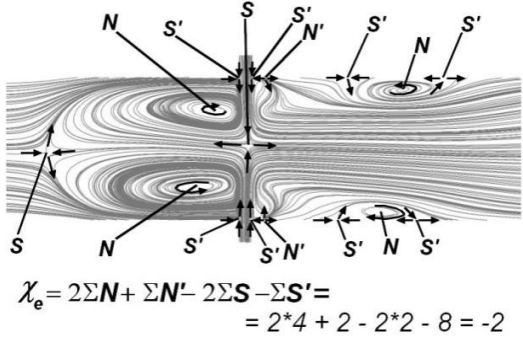
$$\chi_e = 2\Sigma N + \Sigma N' - 2\Sigma S - \Sigma S' =$$
$$= 2*4 + 2 - 2*2 - 8 = -2$$

c) centerplane - jets 2&6

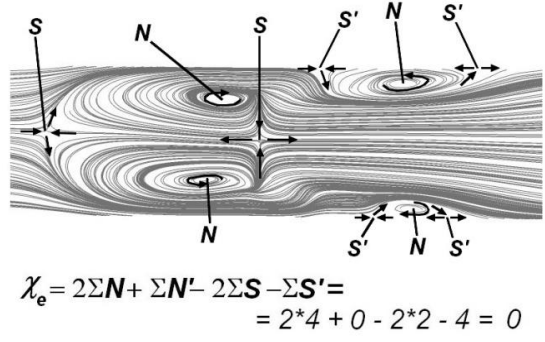
$$\chi_e = 2\Sigma N + \Sigma N' - 2\Sigma S - \Sigma S' =$$
$$= 2*4 + 0 - 2*2 - 4 = 0$$

d) midplane 2.5&6.5

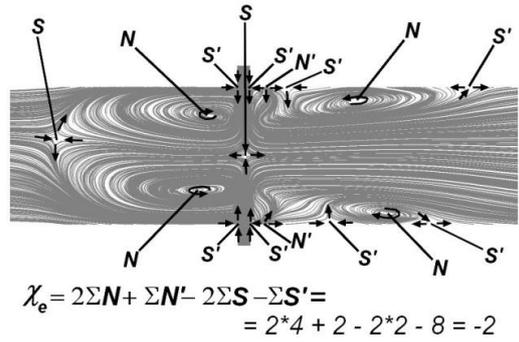
$$\chi_e = 2\Sigma N + \Sigma N' - 2\Sigma S - \Sigma S' =$$
$$= 2*4 + 2 - 2*2 - 8 = -2$$

e) centerplane - jets 3&7

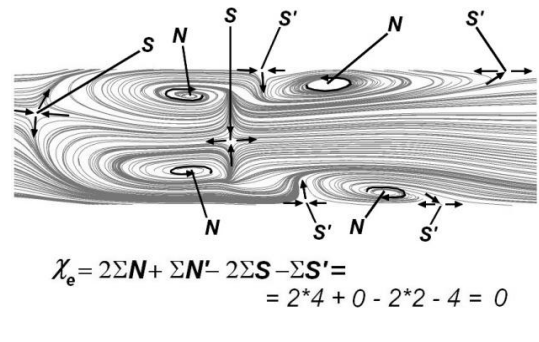
$$\chi_e = 2\Sigma N + \Sigma N' - 2\Sigma S - \Sigma S' =$$
$$= 2*4 + 0 - 2*2 - 4 = 0$$

f) midplane 3.5&7.5

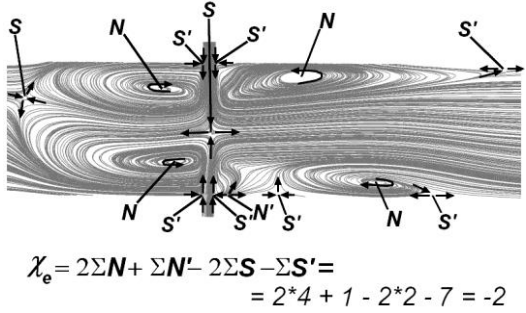
$$\chi_e = 2\Sigma N + \Sigma N' - 2\Sigma S - \Sigma S' =$$
$$= 2*4 + 1 - 2*2 - 7 = -2$$

g) centerplane - jets 4&8

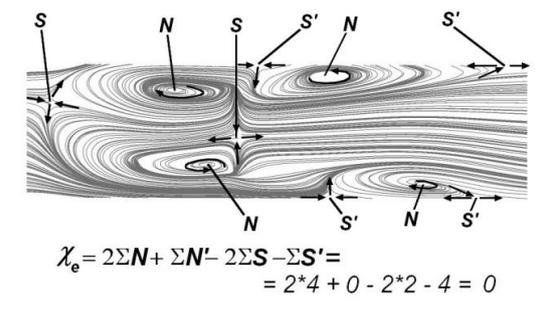
$$\chi_e = 2\Sigma N + \Sigma N' - 2\Sigma S - \Sigma S' =$$
$$= 2*4 + 0 - 2*2 - 4 = 0$$

h) midplane 4.5&8.5

**Fig. 10** *r-x* topology results for *URANS* streamline plots for 360° calculation domain; *n*=8, *DR*=5, $J^{1/2}$ =39.6 (*C*=22.0, *MR*=6.26), *t*=0.730 sec

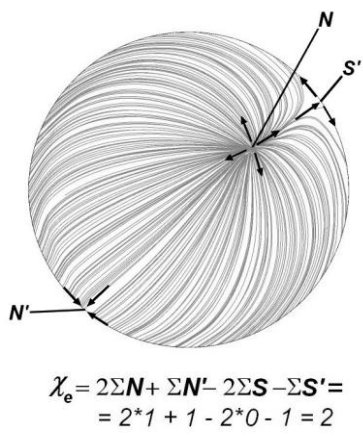
$$\chi_e = 2\Sigma N + \Sigma N' - 2\Sigma S - \Sigma S' =$$
$$= 2*1 + 1 - 2*0 - 1 = 2$$

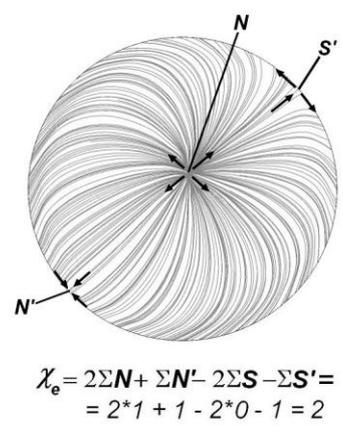
$$\chi_e = 2\Sigma N + \Sigma N' - 2\Sigma S - \Sigma S' =$$
$$= 2*1 + 1 - 2*0 - 1 = 2$$

a) *x/D*= - 1.4                                         b) *x/D*= - 1.0

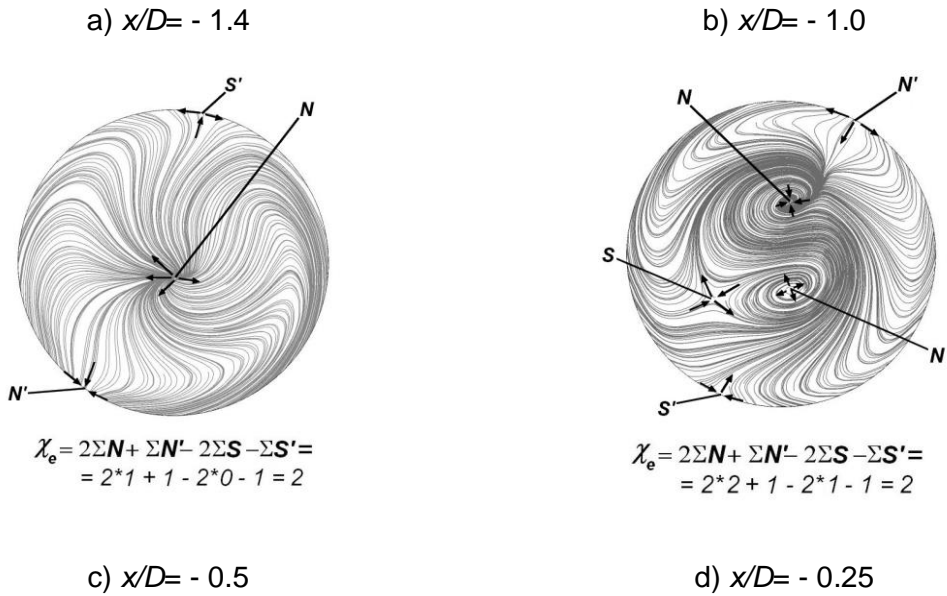

c) *x/D*= - 0.5                                         d) *x/D*= - 0.25

**Fig. 11** *TA* results for *URANS* streamline radial-transverse (*r-θ*) plots for 360° calculation domain; *n*=8, *DR*=5, $J^{1/2}$=39.6 (*C*=22.0, *MR*=6.26), *t*=0.730 sec

The streamline plots in Figs.10&12 are:
1)      centerplane plots (in the left column) for jets 1&5–4&8 for *t*=0.73 &.3.00 sec. respectively;
2)      midplane plots (in the right column) for 1.5&5.5–4.5&8.5 for *t*=0.73 &.3.00 sec. respectively.
     *R-θ* plots for *t*=0.73 & 3 sec. respectively are shown in Figs.11&13. Note that there are 2 nodes and a saddle in 11d which is different from the pattern in 11a-c, but the Rule is still satisfied.

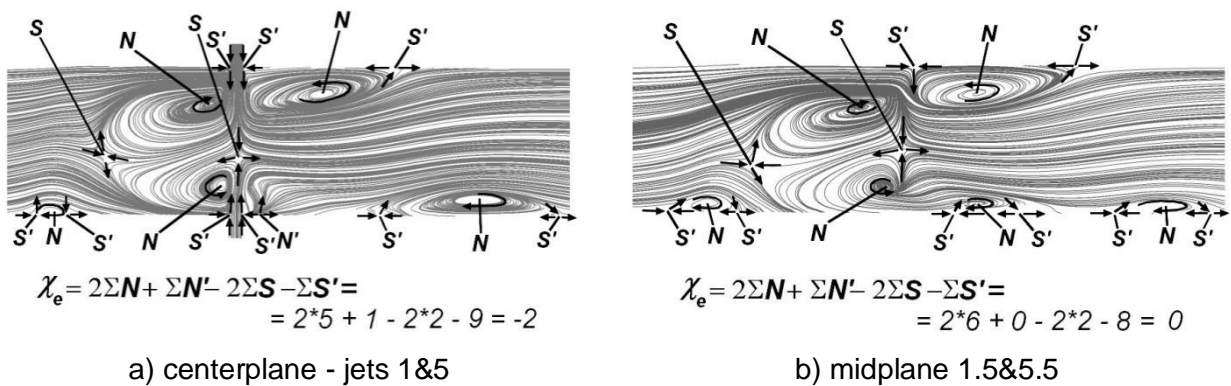

a) centerplane - jets 1&5                               b) midplane 1.5&5.5

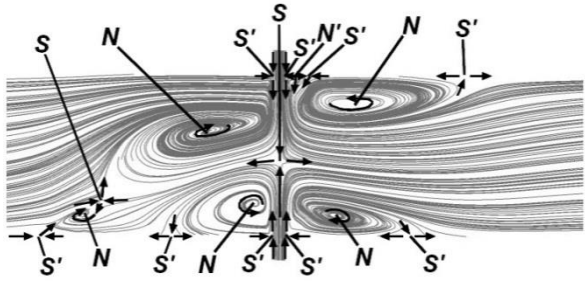

$$\chi_e = 2\Sigma N + \Sigma N' - 2\Sigma S - \Sigma S' =$$
$$= 2*5 + 1 - 2*2 - 9 = -2$$

c) centerplane - jets 2&6

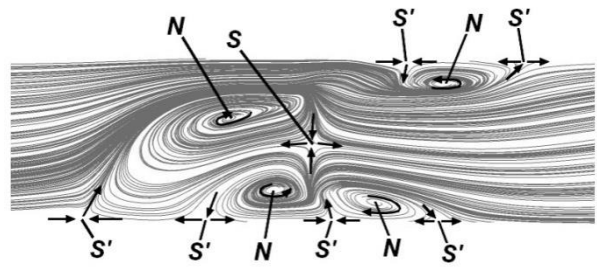

$$\chi_e = 2\Sigma N + \Sigma N' - 2\Sigma S - \Sigma S' =$$
$$= 2*4 + 0 - 2*1 - 6 = 0$$

d) midplane 2.5&6.5

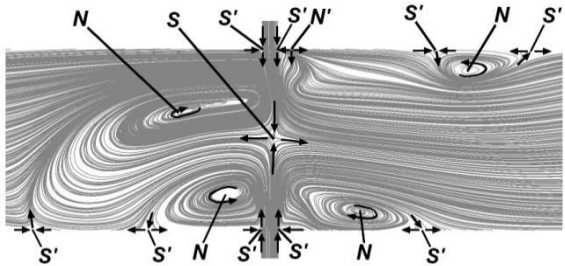

$$\chi_e = 2\Sigma N + \Sigma N' - 2\Sigma S - \Sigma S' =$$
$$= 2*4 + 1 - 2*1 - 9 = -2$$

e) centerplane - jets 3&7

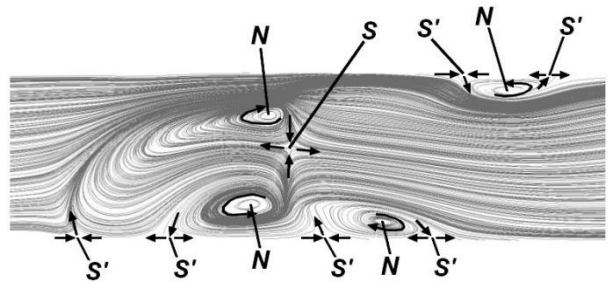

$$\chi_e = 2\Sigma N + \Sigma N' - 2\Sigma S - \Sigma S' =$$
$$= 2*4 + 0 - 2*1 - 6 = 0$$

f) midplane 3.5&7.5

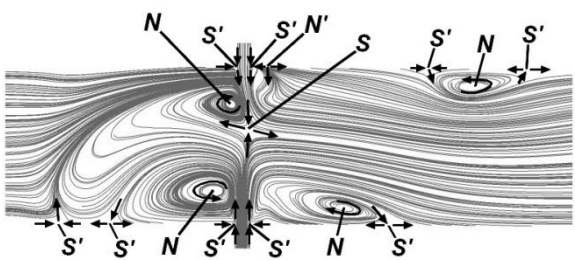

$$\chi_e = 2\Sigma N + \Sigma N' - 2\Sigma S - \Sigma S' =$$
$$= 2*4 + 1 - 2*1 - 9 = -2$$

g) centerplane - jets 4&8

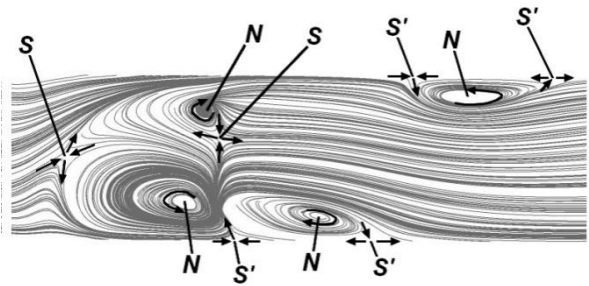

$$\chi_e = 2\Sigma N + \Sigma N' - 2\Sigma S - \Sigma S' =$$
$$= 2*4 + 0 - 2*2 - 4 = 0$$

h) midplane 4.5&8.5

**Fig. 12** *r-x* topology results for *URANS* streamline plots for 360° calculation domain; *n*=8, *DR*=5, $J^{1/2}$=39.6 (*C*=22.0, *MR*=6.26), *t*=3.000 sec

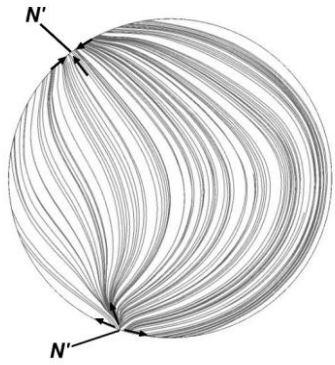

$\chi_e = 2\Sigma N + \Sigma N' - 2\Sigma S - \Sigma S' =$
$= 2*0 + 2 - 2*0 - 0 = 2$

a) $x/D = -1.3$

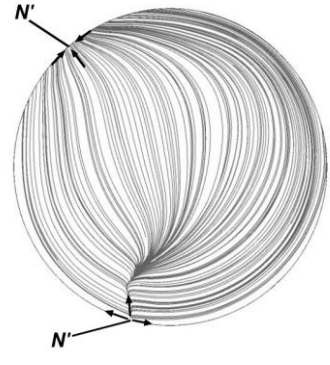

$\chi_e = 2\Sigma N + \Sigma N' - 2\Sigma S - \Sigma S' =$
$= 2*0 + 2 - 2*0 - 0 = 2$

b) $x/D = -1.0$

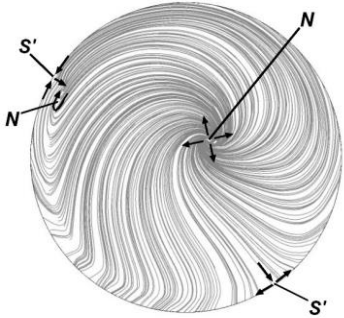

$\chi_e = 2\Sigma N + \Sigma N' - 2\Sigma S - \Sigma S' =$
$= 2*2 + 0 - 2*0 - 2 = 2$

c) $x/D = -0.5$

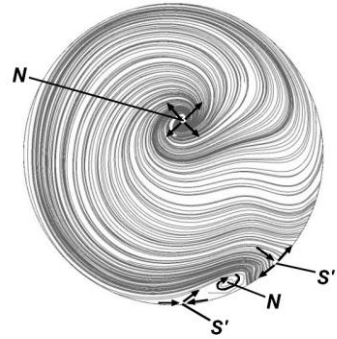

$\chi_e = 2\Sigma N + \Sigma N' - 2\Sigma S - \Sigma S' =$
$= 2*2 + 0 - 2*0 - 2 = 2$

d) $x/D = -0.25$

**Fig. 13** *TA* results for *URANS* streamline radial-transverse ($r$-$\theta$) plots for 360° calculation domain; $n=8$, $DR=5$, $J^{1/2}=39.6$ ($C=22.0$, $MR=6.26$), $t=3.000$ sec

The purpose of *TA* is to verify the integrity of the calculations. Thus to try to assign fluid mechanics reasons to the *TA* results would be 'overthinking' them – except for the obvious ones. Firstly, there are full saddles *S* at *JIP* and the farthest upstream location of the counter-flow jet. Secondly, within all the centerplanes&midplanes in Figs.10&12 there are two pairs of full nodes *N*-at the zero-velocity locations upstream and downstream of *JIP:* the first couple corresponds to the centre of torus-like upstream *RFZ* whereas the second one – torus-like downstream *RFZ*.

## 4 Conclusions

The primary conclusions of this study are as follows. Features of strongly impinging jets in a cylindrical duct are:
- Stagnation points at the farthest upstream penetration of the counter-flow jet and at *JIP*.

- Results were shown for the unsteady counter-flow jet after the commencement of 'flapping'.
- Results were shown at the farthest upstream centerline penetration of the counter-flow jet for strongly impinging jets.
- It was shown that the asymmetry appeared to be cyclic at a frequency of < 2Hz.
- It was shown that the computed results agree with the rules of topological analysis.
- The results for strongly impinging jets revealed large-scale instability of the *JIC* flow field at high *J*'s (*C*'s). The prominent feature of this kind of fluid flow is that two *RFZ*s of continuous torus shape are formed upstream & downstream of jets' injection plane. In general, such a flow configuration would promote an intensive *JIC* flow mixing. These aspects should be accounted in the further analyses of thermal and mechanical loads of components downstream of the mixer.